\documentclass[10pt,conference]{IEEEtran}

\usepackage{cite}
\usepackage{amsmath,amssymb,amsfonts}
\usepackage{algorithm}
\usepackage{algorithmic}
\usepackage{graphicx}
\usepackage{textcomp}
\usepackage{xcolor}
\usepackage{float}
\usepackage{listings}
\usepackage{lstcustom}
\usepackage{hyperref}
\usepackage{tabularx}
\usepackage{silence}
\makeatletter
\newcommand\fs@norules{\def\@fs@cfont{\bfseries}\let\@fs@capt\floatc@ruled
  \def\@fs@pre{}%
  \def\@fs@post{}%
  \def\@fs@mid{\kern3pt}%
  \let\@fs@iftopcapt\iftrue}
\makeatother
\floatstyle{norules}
\restylefloat{algorithm}

\def\BibTeX{{\rm B\kern-.05em{\sc i\kern-.025em b}\kern-.08em
    T\kern-.1667em\lower.7ex\hbox{E}\kern-.125emX}}

\begin{document}

\title{Applying Information Theory to Software Evolution\\
}

\author{\IEEEauthorblockN{Adriano Torres, Sebastian Baltes}
\IEEEauthorblockA{\textit{School of CMS} \\
\textit{University of Adelaide}\\
Adelaide, Australia \\}
\and
\IEEEauthorblockN{Christoph Treude}
\IEEEauthorblockA{\textit{School of CIS} \\
\textit{University of Melbourne}\\
Melbourne, Australia \\}
\and
\IEEEauthorblockN{Markus Wagner}
\IEEEauthorblockA{\textit{Data Science \& AI} \\
\textit{Monash University}\\
Melbourne, Australia \\}

}

\maketitle

\begin{abstract}
Although information theory has found success in disciplines, the literature on its applications to software evolution is limit. We are still missing artifacts that leverage the data and tooling available to measure how the information content of a project can be a proxy for its complexity. In this work, we explore two definitions of entropy, one structural and one textual, and apply it to the historical progression of the commit history of 25 open source projects. We produce evidence that they generally are highly correlated. We also observed that they display weak and unstable correlations with other complexity metrics. Our preliminary investigation of outliers shows an unexpected high frequency of events where there is considerable change in the information content of the project, suggesting that such outliers may inform a definition of surprisal.
\end{abstract}


\section{Introduction}
In a software engineering context, it makes intuitive sense to want to monitor the evolution of entropy, given that managing complexity is a key concept to the construction of software systems~\cite{McConnell:2004:CCS:1096143}. Such a necessity gives rise to several strategies that have become foundational to programming, such as modularity, clean interfaces and design patterns~\cite{gamma1995patterns}.

Software applications that generate value do so by automating and abstracting away complex processes. As software evolves, it is expected for its behaviour to be repaired and augmented, which means the information content of the project invariably changes as the project is maintained. Many projects start by preforming relatively specific tasks, and eventually grow to offer larger sets of functionality in its domain. Consider, for the sake of illustration, that we want to write software that implements the basic operations on real numbers. Upon defining the set of supported operations, one could proceed by incrementally developing the calculator one operation at a time. From an information theoretic perspective, such evolution means the communication channel represented by its source code transmits an increasing amount of information over time.

It common for unmanaged complexity to grow in projects, whether by schedule pressures or by poor design decisions~\cite{mcconnel2018techdebt}. In our calculator, this could occur in different ways; the project could have a single file performing all supported operations in the same file, in which case there would be too much information for a reader to digest at the same time. Given humans' limited working memory capacity, this can potentially lead developers to experience cognitive load~\cite{kalyuga2011cognitive}. Another regular case would be when the code does not leverage natural language properly to convey its meaning, like when one writes code using identifiers that obscure the purpose of the code. If we use the identifier \texttt{`s'} everywhere we want to refer to subtraction, \texttt{`m'} to multiplication and so on, a new contributor will have to spend a lot more time to understand the behaviour of the application. Since the words being used do not convey enough information, the incoming developer will be virtually forced into looking at the implementation of every function being used to understand how the calculator works.

Even though software may be currently behaving properly, it can become difficult for maintainers to extend it due to complexity. Therefore, it is reasonable to speculate that monitoring the evolution of the information content of a project and the units of abstraction that compose it may provide useful insights about its complexity. Examples of such insights could be the generation of warnings when certain files in a system grow so as to contain too much information, or that a newly inserted method appears to be too different from its enclosing scope. Considering the project as a whole, it might be useful to establish certain thresholds of total information transmitted after which it is assumed that there is enough complexity in the system that all changes introduced must be strictly tested.

Given an alphabet $\mathcal{X}$ whose symbols are distributed according to a probabilistic distribution $X$, the Shannon entropy is defined as:

\begin{equation*}
H(X)= -\sum\limits_{x \in X} p(x)* \log_2 {p(x)},
\tag{1}\end{equation*}

Since its formalisation, the concept of entropy~\cite{shannon1948theory} has been applied in many areas of science. Several methods for the calculation of graph entropy~\cite{DEHMER201157history, dehmer2017mathematical} have been explored in different areas of science, with applications spanning through fields like network science~\cite{ephremides1998networks} and chemistry~\cite{Konstantinova2011chemistry, Huang536318}.

The fact that entropy provides a standard unit that does not depend on the system under observation makes it an attractive measure of complexity. Let us examine an elementary case to provide an intuition for why we would prefer less entropy in code. Consider the two snippets of figure \ref{intro-snippet}. From a textual standpoint, it is clear that the first snippet is easier to understand, since it reads almost as plain English. When it provides the entire implementation, the second snippet offers two options to the developer: to read the same algorithmic implementation every time they have to maintain code that uses it, or to ignore the implementation, essentially treating it as noise. In both cases, it seems reasonable to assume that \textit{less explicit information} would be desirable. This is consistent with encapsulation, and reinforces the intuition that, regardless of the unit of abstraction under consideration, it is desirable to manage entropy.

In the real world, the amount of information being hidden behind abstractions that provide us with simple interfaces can be impressive. The actual implementation of Java's \texttt{Math.sqrt}, for instance~\cite{javaSqrtDoc}, contains 115 lines of relatively low level C code~\cite{soJavaSqrt}. Therefore, ensuring that, as software evolves, units of abstraction  - such as files, classes and modules - contain both the right amount and type of information may be useful to the engineering process.

This work aims to provide the following contributions:
\begin{itemize}
    \item A description of the status quo, i.e., previous work done on the applications of information theory as a means of measuring and managing software complexity;
    \item Two definitions of source code entropy: a novel structural definition computed from abstract syntax trees (ASTs), as well as a textual definition, in which the code is regarded as a stream of natural language tokens;
    \item An empirical assessment of the evolution of the different entropy metrics that can be calculated from the two definitions above, using metrics produced from data collected from open source projects, as well an exploration on how such entropy metrics might assist in managing complexity during the software development life cycle.
\end{itemize}

\subsection{Research Questions}
Given that parsers take as input the program's token stream~\cite{Parr10language} to create ASTs, it is natural to wonder how structural and textual entropy curves relate to each other, as well as the extent to which the way the different tokens that are considered influences the measurements. This motivates the following research questions:

\begin{itemize}
    \item \textbf{R.Q.1}: To what extent do textual and structural entropy curves correlate?
    \item \textbf{R.Q.2}: How sensitive to parameters (structural vs textual, tokens considered) are the historical entropy curves of projects?
    \item \textbf{R.Q.3}: How do the entropy metrics correlate to number of lines of code, cyclomatic complexity, and token count?
    \item \textbf{R.Q.4}: How can the evolution of such entropy metrics guide us toward detecting events (commits) in which there is an unusually high variation of the project's total entropy?
\end{itemize}

\begin{figure}
  \begin{lstlisting}[language=Java,numbers=none]
    public double squareRoot(double aPositiveNumber) {
        return Math.sqrt(aPositiveNumber);
    }
  \end{lstlisting}
  \begin{lstlisting}[language=Java, numbers=none]
    public double squareRoot(double aPositiveNumber) {
        double t;
        double squareRoot = aNumber / 2;
        do {
            t = squareRoot;
            squareRoot = (t + (aNumber / t)) / 2;
        } while ((t - squareRoot) >= epsilon);
        return squareRoot;
    }
  \end{lstlisting}
  \caption{Hidden versus explicit information}
  \label{intro-snippet}
\end{figure}

\section{Related Work} \label{literature_review}


Investigations about the usefulness of information theory to measure software complexity date back to the 1970s, when researchers measured the information content structures of that map the program's inputs to its outputs~\cite{hellerman1972lookup}. An entire survey on the topic, which proved foundational to our review, summarised the research on the field from 1972 to 1993~\cite{Khoshgoftaar1994ApplicationsOI}, in which Khoshgoftaar summarises applications of entropy of structures that arise during software specification, design and construction.


\subsection{Code Level} \label{code_level}
At the level of code, preliminary work also dates back to the 1970s, with attempts to partition the code into chunks of statements that are guaranteed to keep the control flow inside the sequence of statments of each chunk, deriving an entropy measure for control flow graphs that represent the dependencies between the chunks, as well as their content and size~\cite{davi88complexity}. Other researchers also approached the task of calculating the entropy of control flow structures that represented the nesting between \texttt{if} statements and measured the impact of the entropy of such graphs in developer productivity~\cite{chen1978complexity},~\cite{khoshgoftaar1998information}. However, their measurement of productivity involved close monitoring of developers' production process. Not only does this raise practical concerns, but ethics and privacy matters may also impose a barrier to the maturing of this kind of research.

Other researchers proposed calculations for source code entropy that relied on word-based metrics extracted from compiler tokens, wherein source code is seen as a distribution of names and symbols. Although they do not provide a method to create an artifact that meets their requirements, and consequently, did not provide any empirical validation of their method, Berlinger states that a programming environment that satisfies the following properties could provide enough infrastructure for entropy measurements~\cite{berlinger1980complexity}:
\begin{itemize}
    \item Statistics on the percentage of operators, parentheses, and other code tokens are known;
    \item Ability to rank probabilities of operands such as arrays, constants, function names by frequency of use;
    \item Frequencies for labels are also maintained;
    \item Ability to keep track of structures that create nested control flow structures, such as \texttt{DO} and \texttt{WHILE}
\end{itemize}

Berlinger argues that such conditions would enable the creation of a fully automated process that can collect such metrics early on during development, thus presenting a potential way to manage code complexity since the beginning of the project. The main limitation from the his point of view for such a system to be viable would be the ability to calculate the distributions of the tokens, given the constraints in code and text processing capacity at that time. Other explorers of this venue also worked under unrealistic grounds for the token distributions, the most common being the assumption that all operators, operands or other atomic units of measure were produced by an equiprobable stochastic process~\cite{halstead1977elements}. 
There is also evidence that information theoretic metrics can be a better proxy for predicting whether a module is difficult to work with than number of lines of code or cyclomatic complexity ~\cite{cook1993information}, as per McCabe's definition of it being equal to the maximum possible number of linearly independent cycles in the program's Control Flow Graph ~\cite{mccabe1976complexity}. An interesting observation they arrived to was that even though the number of instructions per file varied by a factor of over 100, the distribution of instructions or their classes was mostly consistent. They also conclude by remarking that ``rarely used instructions are a major contributor to the difficulty in understanding [assembly] programs'', and that the detection of such \textit{surprising} instructions may ``aid in allocating program maintenance resources and in pointing out areas of potential difficulty''~\cite{cook1993information}. Further developments in the field involved measurements of synthetic measures in which correlation matrices produce distributions whose entropy would serve as a proxy for overall program complexity~\cite{khoshgoftaar1992measure}. The same researcher also proposed a complexity definition based on the Kolmogorov complexity~\cite{kolmogorov1965three} of a set of metrics computed to estimate module complexity. Ultimately, the calculation of the Kolmogorov complexity of such a structure is not computable, and therefore their results provide only approximations of a measure which, in turn, is to be taken as a proxy of the overall complexity.

The most recent developments in the field stem from empirical assessments about the evolution of the entropy of the files that compose the project, having the number of lines changed at arbitrary time slices as a basis for measuring entropy, and calculating individual entropy for each file, relative to the entire context of the project. Hassan ~\cite{hassan2009predicting} defends the intuition that software systems that are changed in several different locations cause developers to have difficulty in keeping track of all the changes, therefore attributing a higher entropy to scattered changes~\cite{hassan2009predicting},~\cite{canfora2014changes}. Though keeping track of such changes might indeed prove harder for code reviewers, sacrificing modularity and encapsulation might be a compromise not worth undertaking. Evidence produced by different groups of researchers who built on top of this definition can be conflicting, as some authors found that refactoring has beneficial effects towards reducing entropy~\cite{canfora2014changes}, whereas other results suggested that trends in entropy are irrespective of refactoring activity~\cite{keenan2022investigation}. All of these recent publications employ modern data and code mining capabilities, and all of them perform small-scale empirical evaluations of their hypotheses. While this, relative to previous studies, a clear advancement, we believe that metrics of entropy that are based on the number of lines changed - ultimately metadata about the change - might be missing out on information that is represented by the code itself.

Our survey of the state of the art suggests that questions that were raised over 20 years ago, and are still relevant to the software engineering community, remain largely unaddressed~\cite{Khoshgoftaar1994ApplicationsOI}:
\begin{itemize}
    \item What empirical relationships can be established from large scale studies involving industrial size software systems?
    \item Which information theory measures represent dimensions of complexity measures that are not captured by more conventional software metrics, such as lines of code, cyclomatic complexity and token count?
    \item Is there enough evidence to back up the intuition that the more information a software project contains, the harder it appears to software engineers?
\end{itemize}

The literature we found on the topic leads us to the conclusion that further explorations of the information content of ways of representing source code entropy may present promising research venues. For example, although much of the published material in the field revolves around graph entropy metrics, we found no information theoretic studies regarding the information content of abstract syntax trees, and how they correlate to the entropy of data/flow control graphs, or code chunks. 
We also believe that, given the unprecedented availability of both open source code, as well as software and hardware that enables large scale analysis, token-based analysis can also provide actionable insights. In fact, there is a growing number of recent publications concerned with representations of code for machine learning engines, which employ token-based analysis to tasks like semantic labeling of code snippets~\cite{allamanis2016convolutional}, as well as for predicting code comments~\cite{movshovitz-attias-cohen-2013-natural}.

\section{Definition of Entropy} \label{definition}
In this work, we analyse the entropy of a project from two main standpoints: structural - considering the edges of the AST that represents the code - and textual, wherein natural language tokens are extracted from the source code.

Intuitively, one would expect that the entropy of the AST edges would be able to detect when two pieces of code are similar or different in structure. In our calculator example, it is reasonable to assume that the structural entropy will be able to determine that methods that add and multiply numbers are very similar in structure, i.e., they both take two parameters, perform an operation using the two input values, and produce an output type. However, AST's do not contain identifier names, as well as some other symbols that are resolved during semantic analysis, which means that symbols like \texttt{addNumbers} or \texttt{multiplyNumbers} are not available to the parse tree. If, for example, our calculator contains only operations on numbers, if a newly introduced method is named \texttt{splitStrings} or \texttt{requestHttpServer}, textual entropy should be able to detect that this method is out of context. In this work, we restrict ourselves to file contexts, as files are natural units of compilation. Wider contexts, such as the entire project at a given point in time, or ``global'' contexts, which merges contexts from multiple projects, shall be explored in future research.

Let $\mathcal{P}$ be a project that contains a set $\mathcal{F}$  of source code files, which are known to be syntactically correct. Let also $\mathcal{X}_{f}$ represent the distribution of the frequencies of the edges that form the AST of file $f$, represented by $AST(f) = (N, E)$, where $N$ is a set of nodes, and $E$ is a set of edges. We define a \textit{context} $\mathcal{C} = \bigcup_{f\in{S}}\mathcal{X}_{f}$, where $S\subseteq F$ is a set of files, representing a distribution of all the AST edges of the respective context. We define \textit{structural entropy} as:

\begin{equation*}
H_{AST}(file\,|\,\mathcal{C})= -\sum\limits_{edge \in E} p(edge\,|\,\mathcal{C})* \log_2 {p(edge\,|\,\mathcal{C})},
\tag{2}\end{equation*}

where

\begin{equation*}
{p(e\,|\,\mathcal{C})=
\begin{cases}
p(e\,|\,\mathcal{X}_{f}), & \mathcal{C} = \mathcal{X}_{f},\\
\frac{p(e\,|\,\mathcal{X}_{f})}{p(e\,|\,\mathcal{C})},& \text{otherwise.}
\end{cases}}
\end{equation*}

In this scenario, the AST edges are calculated by first parsing the source code and producing a parse tree. Breadth-first traversal is then applied, and each pair of parent and child nodes constitutes and AST edge that is kept in a histogram.

When the context is restricted to the file itself, the entropy is calculated based on the relative frequencies of the edges within the file, whereas, for larger contexts, the entropy considers the relative frequencies between the two contexts.

It follows naturally from the logarithmic definition of entropy that the entropy of two separate files correspond to the sum of the separate entropies of the file. Therefore, such a definition allows us to additively compute it for the entire project at any point in time.

The definition of \textit{textual entropy} of the file is analogous, but instead we consider the set $\mathcal{W}$ of words that compose the file:

\begin{equation*}
H_{TOKEN}(file\,|\,\mathcal{C})=-\sum\limits_{word \in W} p(word\,|\,\mathcal{C})* \log_2 {p(word\,|\,\mathcal{C})},
\tag{3}\end{equation*}

Although Shannon's definition of entropy has been applied in different ways to measure code complexity, we found no work in the literature that attempted to measure structural entropy from the AST edges; the papers surveyed in section \ref{literature_review} concerned themselves only with graph entropy measures, either from Control Flow Graphs, or from high level specification graphs. As to token entropy, even though we found applications of the same formula we are applying, they restricted themselves to low level assembly code \cite{cook1993information}. Moreover, in this work, we make no assumptions about the distribution of the tokens. Instead, we leverage the vast amount of data available in open source projects to directly calculate their distributions for each project that was analysed. Finally, we make measurements about the actual data of the code, as opposed to its metadata, and do not apply arbitrary time slicing, i.e., we consider all commits in the history of .

\subsection{Verifying The Definition In A Controlled Environment}
In order to perform an initial validation of this definition in a simple and controlled environment, we created a software repository and started a basic implementation of a calculator. We progressively added code (i.e., AST edges and tokens) that performs addition, subtraction and multiplication, measuring the net entropy of each commit. Other experiments we conducted involved comparing different implementations that were expected to yield the same result (e.g., figure \ref{intro-snippet}). Once there was enough information in the file to not make it overly sensitive to any change, we started introducing unrelated functionality, mostly in the form of string operations, which were ultimately extracted out from the main calculator file and into its own context. A total of 34 commits were introduced, and we labeled our expectations using the following encoding: \textbf{-1}: Entropy decrease expected; \textbf{0}: No change expected \textbf{1}: Entropy increase expected. Upon labelling all commits, we applied our definitions of structural and textual entropy, varying the tokenisation method by progressively removing Java keywords and then numbers. As an example of our tokenisation method, consider a function whose signature is \texttt{public String decodeStreamFromBase64ToBase32(String message)}. Its succesive tokenisation steps would result in the following sets of words:
\begin{itemize}
    \item Basic: \texttt{public}, \texttt{string}, \texttt{decode}, \texttt{stream}, \texttt{from}, \texttt{base}, \texttt{64}, \texttt{to}, \texttt{base}, \texttt{32}, \texttt{string}, \texttt{message};
    \item Keywords removed: \texttt{public} and the two occurences of \texttt{string} are discarded;
    \item Keywords and numbers removed: \texttt{32} and \texttt{64} are removed.
\end{itemize}

We then measured the Spearman correlations between the entropy deltas and our expectations. For the 34 data points, we observed a priori correlation of ~0.79. Upon manually checking the data and performing step-by-step analysis, we arrived at a correlation of ~0.98, i.e., only one measurement where the data did not match our expectation. Such a point represented a 0.03\% increase in entropy due to the removal of a redundant AST edge, which is probably too small for human reasoning to seize, but was still captured by our definition.

\section{Study Settings}
In this section, we describe how we applied the two definitions established in section \ref{definition} to calculate the historical evolution of the entropy of open source projects. In our context, historical entropy is defined by the plot derived from the total sum of the entropies of all source code files in the project, measured successively, commit by commit.

\subsection{Project Selection Criteria}
Due to practical matters, we restricted the scope of this exploration to Java repositories publicly available on GitHub. Our decision in both cases is closely related to the availability of open source tooling; we used JavaParser to process code targeted for multiple versions of Java~\cite{javaparser}, and we used the GitHubSearch API \cite{Dabic:msr2021githubSearch} to query for repositories that matched our criteria.

Given the number of repositories available on the platform, and its public nature, it becomes difficult for researchers to determine which projects are actively maintained~\cite{munaiah2017curating}. Though the GitHubSearch API already filters out repositories with less than 10 stargazers, unfiltered queries for Java projects can return as many as 88000 entries. In order to filter this list to find the most active and widely used projects, our query to the engine included the following parameters:

\begin{itemize}
    \item Activity: Project has license, open pull requests, and open issues. The last commit to the repository is less than six months old;
    \item Stargazers count is in the 99th percentile;
    \item Commit count is in the 95th percentile;
\end{itemize}

This resulted in a dataset with 65 projects, from different application domains (distributed applications, web frameworks, virtual machines, security libraries, database managers), many of them widely used. Each project's repository was manually visited, and some projects were removed; common reasons being a) the project was based in languages that used an alphabet that is different from English's (e.g., Chinese and Russian), and b) the project was not a production application. Some projects were also removed because the libraries we used to process the commit history crashed internally whilst executing. This narrowed the list down to 41 entries, from which we selected 25 uniformly at random. The decision to select 25 projects was primarily informed by computational constraints, given the experiments were run on a single core. A detailed summary of the candidate projects selected, as well as the reasons why some of them included are provided in the supplementary material. The engineering and design of solution that leverages parallelism, and that provides better robutstness and fault tolerance, thus being able to circumvent the limitations of the crashes caused by the auxiliary libraries utilized to process the projects, is topic of our future research.

\subsection{Methods}
Using the PyDriller~\cite{PyDriller} open source utility, we traversed all the commits of each project in historical order. Each file's code string served as input to two workflows: a JavaParser~\cite{javaparser} subprocess that parsed the code into the ASTs and generated the histogram of the edges, and a Python library that extracts natural language tokens from source code~\cite{Hucka2018spiral}. In order to collect as much data as possible, all \texttt{.java} files present in the project were considered. The contexts of all the modified files in each commit were processed both as ASTs and as token streams, and their net entropies were measured. For each file, the entropy of the edges of the AST were calculated, and textual entropy was measured in three ways: full text, without Java keywords and without keywords and numbers. For each such metric, their normalised version - the net entropy contrubution divided by the maximum entropy possible, which happens when all tokens are equiprobable - was also calculating, generating a total of eight metrics.
To answer \textbf{R.Q.1} and \textbf{R.Q.2}, we created discrete plots whose horizontal axis represents the commit count - accumulated in chronological order - and the vertical axis measures total entropy in bits. We also created correlation tables between the eight aforementioned entropy metrics and compared them across the projects.
During the processing of the commits, we also keep track of the number of modified lines during the commit, as well as the number of modified tokens and cyclomatic complexity. The correlation between these metrics and the different entropy distributions will provide insights into \textbf{R.Q.3}.

For \textbf{R.Q.4}, our main goal is to investigate how the concept of entropy can help us define and quantify surprise in the context of software evolution. As an initial exploration, we are going to count the number of events whose net contribution to the entropy was an outlier in their respective distributions. These conservative measurements shall contribute to a more complete definition of surprisal, which we will develop in further research.

\section{Empirical Assessment}
This section follows from our analysis of the data generated for the 25 projects selected in the experiment. The code that produces the raw data, some intermediate versions, and a superset of the pictures herein presented can be found it \cite{anonymousRepo}.
The data mining process was conducted on a local, 4-core, 8-thread Intel(R) Core(TM) i7-7700HQ CPU @ 2.80GHz machine, with a total of 32GB available. Three processes at a time were run concurrently. Median processing time was 4.66 hours, with the fastest and slowest runs taking, respectively 0.89 and 114.3 hours.

\subsection{Evolution Of Entropy And Their Correlations}

\begin{figure}
    \centering
    \includegraphics[width=\linewidth,trim=0 30 60 90,clip]{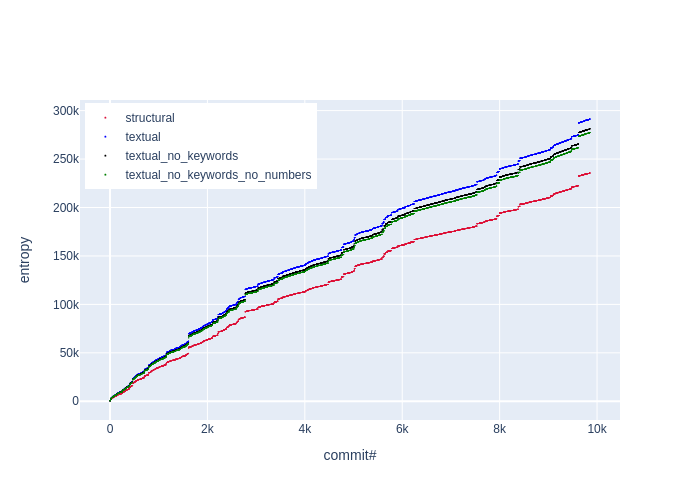}
    \caption{Evolution of file-level entropy aggregated by commit for \texttt{libgdx}, a game development framework.}
    \label{history-libgdx}
\end{figure}

\begin{figure}
    \centering
    \includegraphics[width=\linewidth,trim=0 30 60 90,clip]{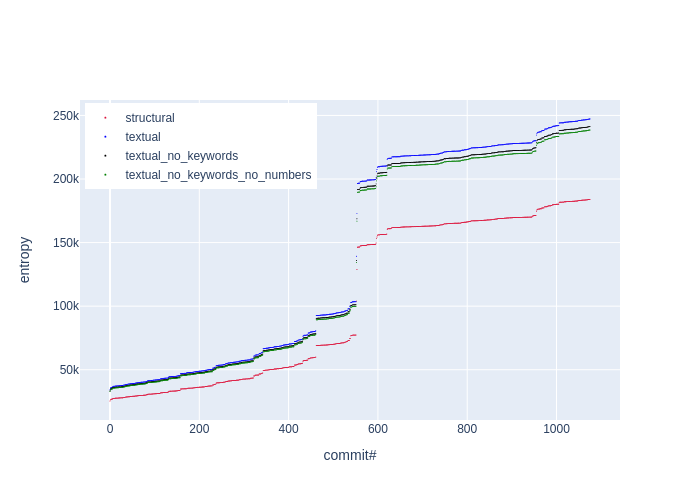}
    \caption{Entropy curve of \texttt{deeplearning4j}, showing big increases in the entropy.}
    \label{history-dl4j}
\end{figure}

Figure~\ref{history-libgdx} shows the typical evolution of each way of calculating project entropy according to our definitions. All projects analysed displayed increasing entropy over time, and all projects exhibited the presence of spikes that were detectable by looking at the plots. Figure~\ref{history-dl4j} shows an example of a project that contains commits where there is considerable variation of entropy.

\begin{figure}
    \centering
    \includegraphics[width=\linewidth,trim=0 30 60 90,clip]{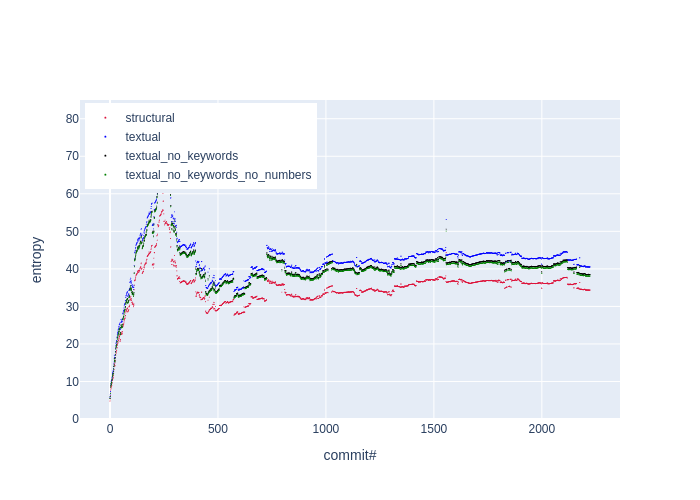}
    \caption{Entropy per number of Java files for \texttt{lombok}, which extends the Java language.}
    \label{history-lombok-per-file}
\end{figure}

When the total entropy is divided by the number of Java files in the repository, projects display less stable trends. On one extreme is the project listed in Figure~\ref{history-dl4j}, whose curve remains almost the same when file count is considered, whereas Figure~\ref{history-lombok-per-file} shows a project whose initial entropy per file initially increased rapidly, but stabilises as the project evolves. This can be a sign of refactoring efforts to maintain complexity under control.

The aforementioned graphs are in accordance with Lehman's first two laws of program evolution, namely \cite{lehman1980laws}:
\begin{itemize}
    \item Continuing Change: all of the sampled projects generate value in their respective domain, and as such, are required to constantly change to either add and correct functionality, or to fix incorrect behaviour;
    \item Increasing Complexity: The upward trends in both structural and entropy metrics reveal that complexity as a proxy for deteriorating structure tends to increase in the projects. Still, some projects - such as \texttt{lombok} and \texttt{grpc} - display trends that make it clear that effort was made to keep complexity under control, whereas others, from an entropy standpoint, displayed uncontrolled, growing complexity.
\end{itemize}

The correlation heatmaps of the 8 entropy curves are shown in Figure~\ref{fig:entropy-correlations}. In the picture, correlations that are close to zero are depicted in white, with darker shades of green represeting values that are close to 1. The majority of the correlations were situated between 0.7 and 0.9. As we can see, all but two projects display strong correlations between all computed metrics, two exceptions being \texttt{jedis} and \texttt{lombok}, which showed low correlation between structural and textual entropy, regardless of the tokenisation method. Manual inspection of the projects revealed that both codebases consist primarily of interfacing/wrapping code, therefore producing ASTs that are relatively simple and reduntant. Both projects also display high  amounts of natural language comments relative to lines of code, which is not unusual for code that provides high level abstractions or APIs. We applied the Dancey and Reidy categorization of correlation coefficients \cite{akoglu2018coefficients}, in which correlations with absolute value between 0.1 and 0.3 are weak, 0.4 to 0.6 are moderate, and 0.7 to 0.9 are strong (zero being none, and one being perfect). We arrive at the following findings for \textbf{R.Q.1} and \textbf{R.Q.2}:
\begin{itemize}
    \item All projects show an increase in entropy over time. 84\% (N = 21) projects display a similar pattern when the file count is considered: very fast initial growth, followed by a stabilisation period. Only 4 projects did not attempt to keep the average entropy per file under control;
    \item 68\% of the projects display strong correlation between structural and textual entropy; the remaining projects display moderate correlations. For two projects, structural and textual entropy correlate highly, but only moderately as the tokenisation method becomes less accepting;
    \item Structural entropy is consistently lower than textual entropy. The distributions for the different tokenisation methods are almost perfect in all the projects, suggesting that language keywords and numbers do not affect the overall information content significantly.
\end{itemize}

\def \heatmapWidth {16mm}

\begin{figure}[htb!]
\centering
\centering
 \includegraphics[width=\heatmapWidth, keepaspectratio]{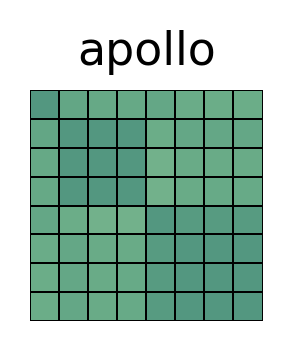}
 \includegraphics[width=\heatmapWidth, keepaspectratio]{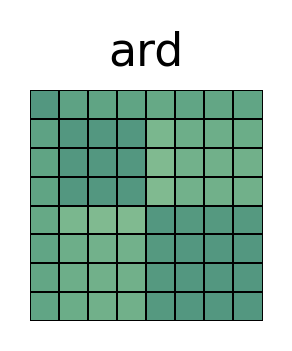}
 \includegraphics[width=\heatmapWidth, keepaspectratio]{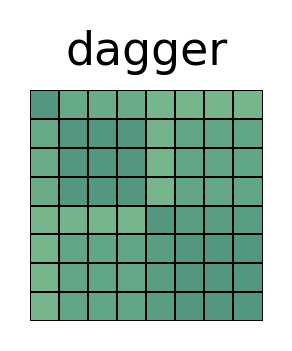}
 \includegraphics[width=\heatmapWidth, keepaspectratio]{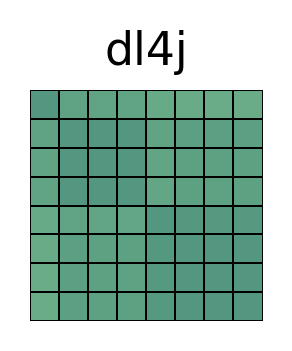}
 \includegraphics[width=\heatmapWidth, keepaspectratio]{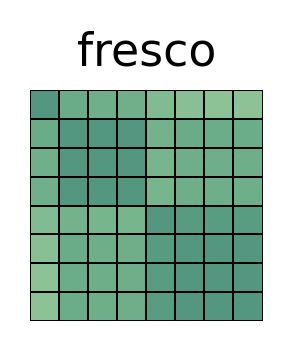} \\
 \includegraphics[width=\heatmapWidth, keepaspectratio]{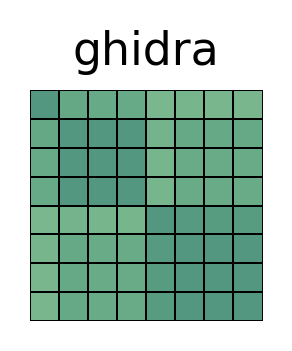}
 \includegraphics[width=\heatmapWidth, keepaspectratio]{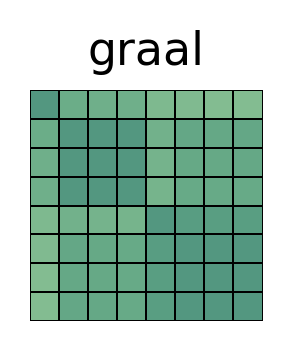}
 \includegraphics[width=\heatmapWidth, keepaspectratio]{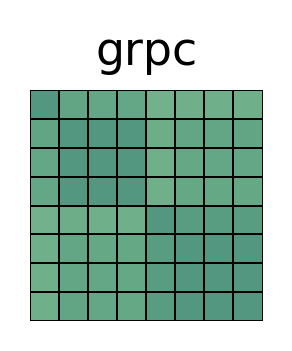}
 \includegraphics[width=\heatmapWidth, keepaspectratio]{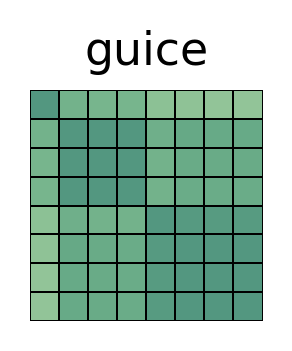}
 \includegraphics[width=\heatmapWidth, keepaspectratio]{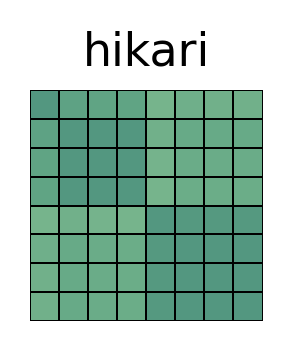} \\
 \includegraphics[width=\heatmapWidth, keepaspectratio]{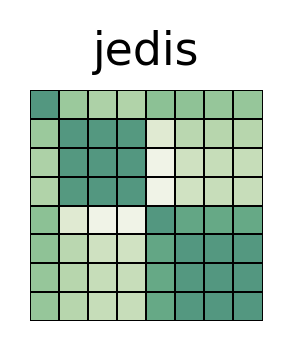}
 \includegraphics[width=\heatmapWidth, keepaspectratio]{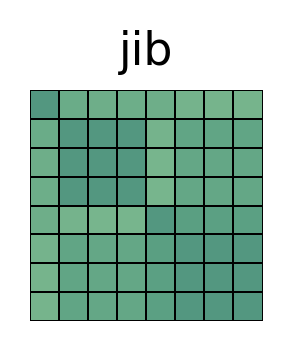}
 \includegraphics[width=\heatmapWidth, keepaspectratio]{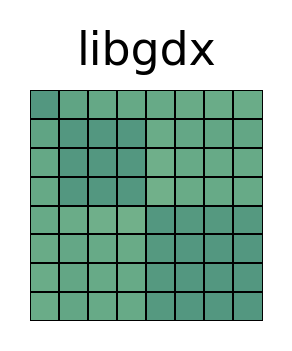}
 \includegraphics[width=\heatmapWidth, keepaspectratio]{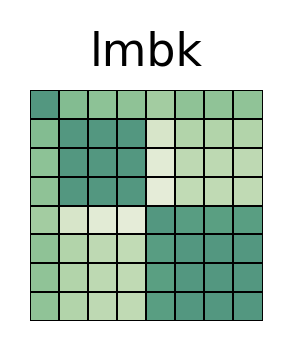}
 \includegraphics[width=\heatmapWidth, keepaspectratio]{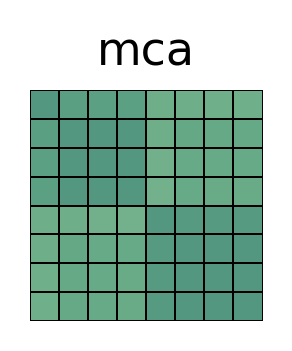} \\
 \includegraphics[width=\heatmapWidth, keepaspectratio]{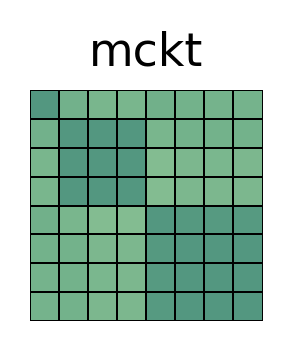}
 \includegraphics[width=\heatmapWidth, keepaspectratio]{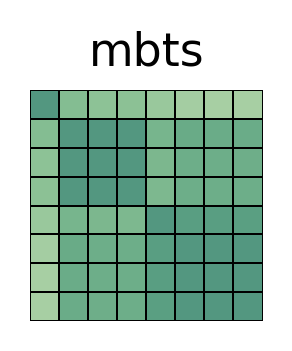}
 \includegraphics[width=\heatmapWidth, keepaspectratio]{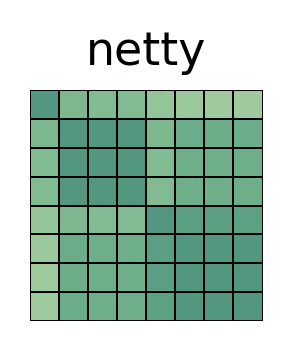}
 \includegraphics[width=\heatmapWidth, keepaspectratio]{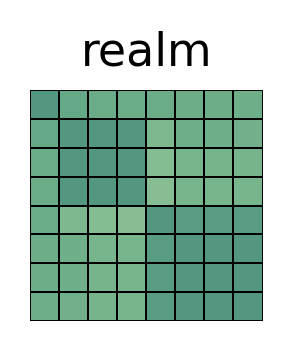}
 \includegraphics[width=\heatmapWidth, keepaspectratio]{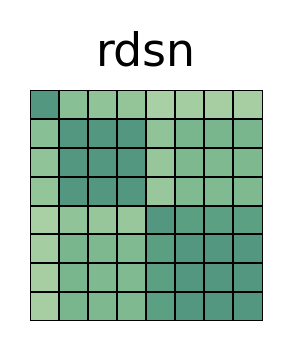} \\
 \includegraphics[width=\heatmapWidth, keepaspectratio]{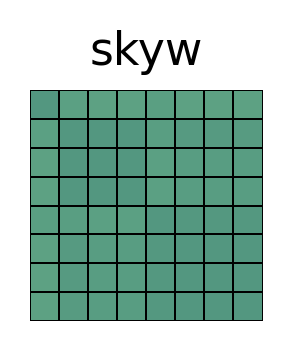}
 \includegraphics[width=\heatmapWidth, keepaspectratio]{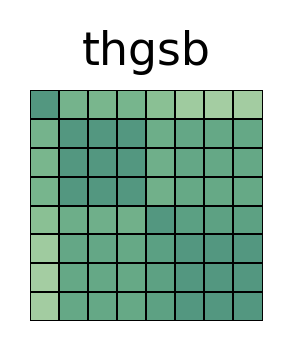}
 \includegraphics[width=\heatmapWidth, keepaspectratio]{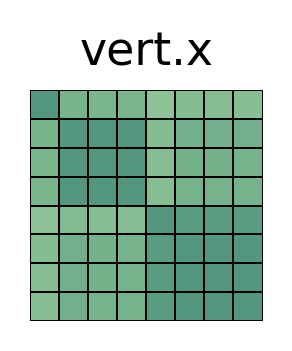}
 \includegraphics[width=\heatmapWidth, keepaspectratio]{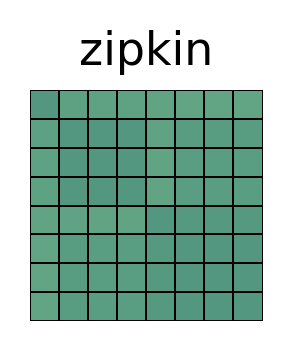}
 \includegraphics[width=\heatmapWidth, keepaspectratio]{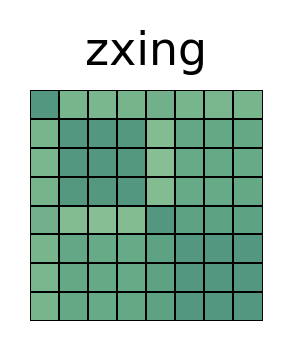} 
\caption{Correlations between structural and textual entropy, with and without normalisation. The darker the square, the closer the value is to 1. White values represent 0.}
\label{fig:entropy-correlations}
\end{figure}

\subsection{Correlation With Standard Software Metrics}
In this section we present our findings regarding \textbf{R.Q.3}.
We measured the correlation between our four entropy metrics (one structural, three textual) with the number of modified lines, the number of modified tokens and cyclomatic complexity of the resulting file. The results are displayed in figure \ref{fig:classic-correlations}. Notice that the correlations are now much weaker, ranging from 0 to no more than 0.54 with some of them approaching zero. We also observed that, in 23 (92\%) projects, textual entropy has stronger correlations with the classic than structural entropy does. This makes intuitive sense, given that tokens are the actual constituents of the lines of code, and that one of the metrics is exactly token count. Correlations with structural entropy are mostly weak, and moderate at best. 12 projects displayed weak correlation for textual entropy, whilst another 12 had a moderate correlation. A single project showed strong correlation. For several projects, the tokenisation method was responsible for bringing the correlation from moderate to low.

\def \heatmapWidth {16mm}
\begin{figure}[htb!]
\centering
\centering
 \includegraphics[width=\heatmapWidth, keepaspectratio]{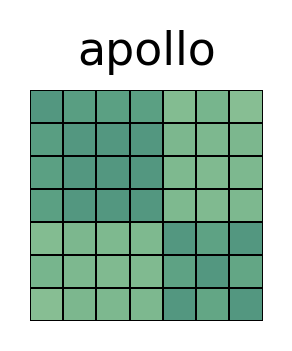}
 \includegraphics[width=\heatmapWidth, keepaspectratio]{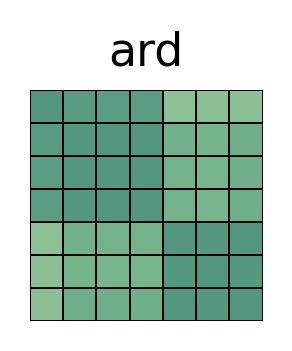}
 \includegraphics[width=\heatmapWidth, keepaspectratio]{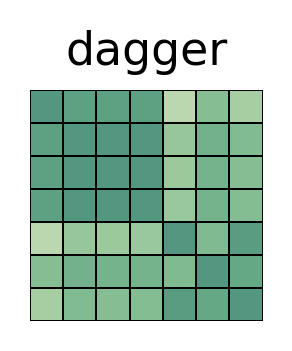}
 \includegraphics[width=\heatmapWidth, keepaspectratio]{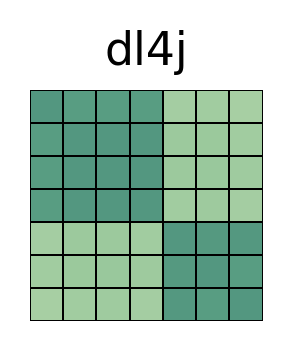}
 \includegraphics[width=\heatmapWidth, keepaspectratio]{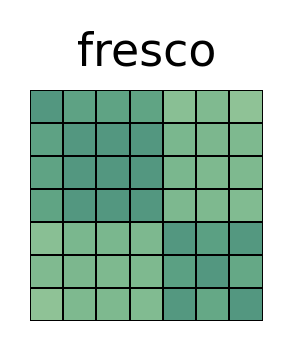} \\
 \includegraphics[width=\heatmapWidth, keepaspectratio]{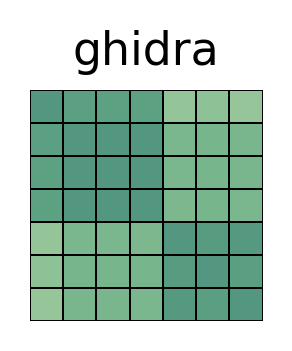}
 \includegraphics[width=\heatmapWidth, keepaspectratio]{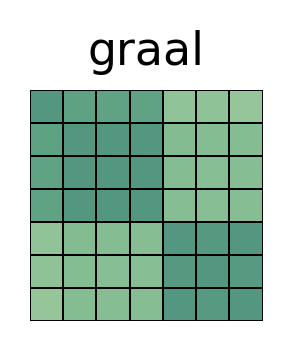}
 \includegraphics[width=\heatmapWidth, keepaspectratio]{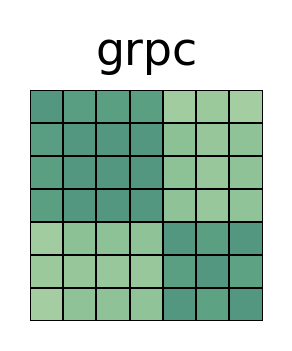}
 \includegraphics[width=\heatmapWidth, keepaspectratio]{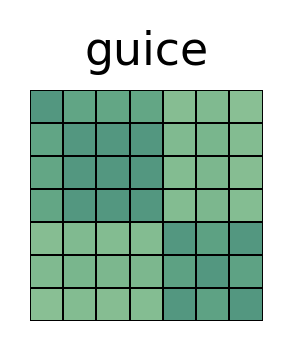}
 \includegraphics[width=\heatmapWidth, keepaspectratio]{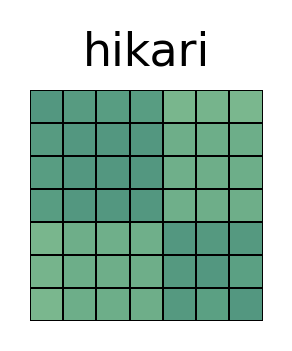} \\
 \includegraphics[width=\heatmapWidth, keepaspectratio]{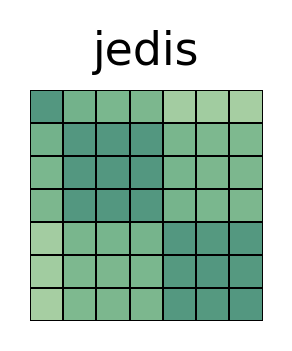}
 \includegraphics[width=\heatmapWidth, keepaspectratio]{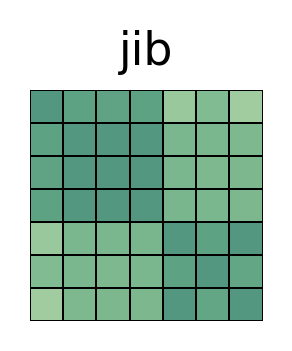}
 \includegraphics[width=\heatmapWidth, keepaspectratio]{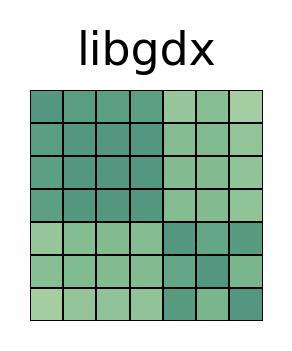}
 \includegraphics[width=\heatmapWidth, keepaspectratio]{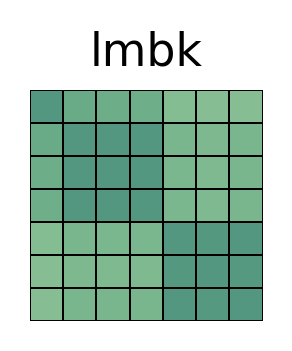}
 \includegraphics[width=\heatmapWidth, keepaspectratio]{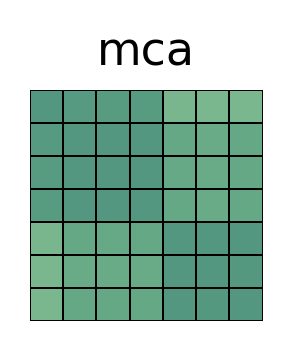} \\
 \includegraphics[width=\heatmapWidth, keepaspectratio]{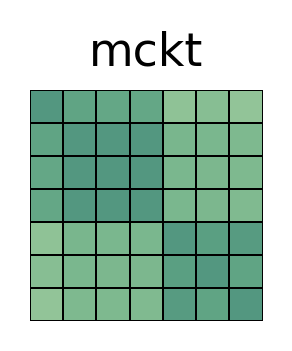}
 \includegraphics[width=\heatmapWidth, keepaspectratio]{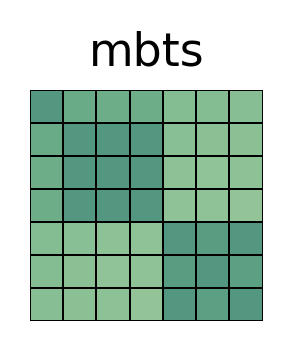}
 \includegraphics[width=\heatmapWidth, keepaspectratio]{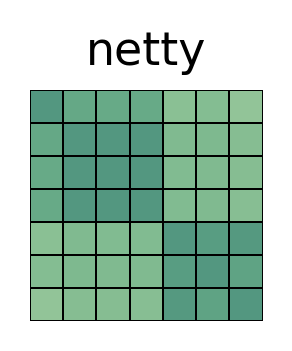}
 \includegraphics[width=\heatmapWidth, keepaspectratio]{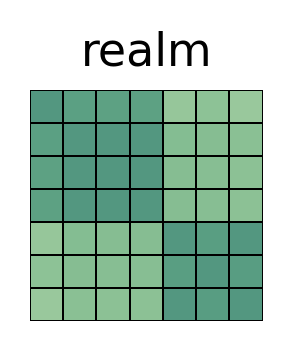}
 \includegraphics[width=\heatmapWidth, keepaspectratio]{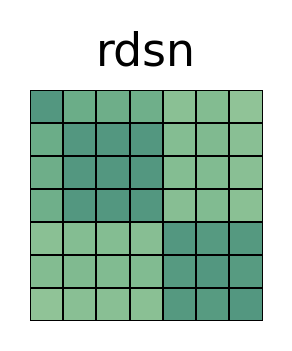} \\
 \includegraphics[width=\heatmapWidth, keepaspectratio]{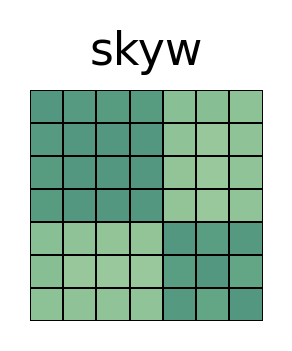}
 \includegraphics[width=\heatmapWidth, keepaspectratio]{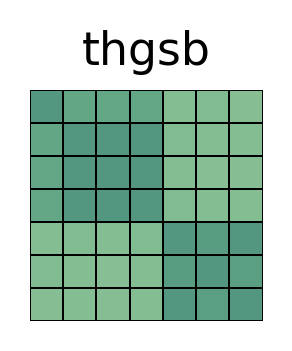}
 \includegraphics[width=\heatmapWidth, keepaspectratio]{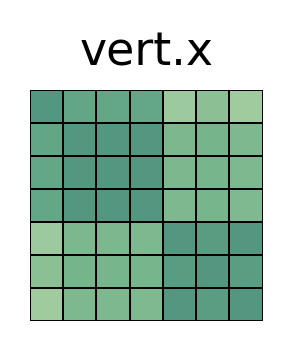}
 \includegraphics[width=\heatmapWidth, keepaspectratio]{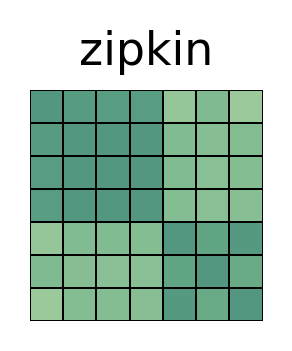}
 \includegraphics[width=\heatmapWidth, keepaspectratio]{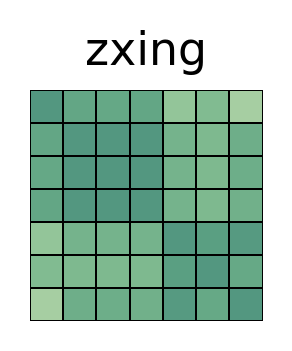} 
\caption{Correlations between entropy metrics and classic metrics. Darker values represent high positive correlation, whereas white means high negative correlation.}
\label{fig:classic-correlations}
\end{figure}

\subsection{Outliers In Entropy Variation}
In order to build an intuition for surprisal in a software engineering scenario, we calculated the outliers for structural entropy and the successive removal of natural language tokens, as well as their normalised counterparts. The complete tables are omitted for brevity, and are provided in the supplementary material \cite{anonymousRepo}.
Using a 1.5 times the inter-quartile range to detect outliers, the projects averaged 12\% of cases for structural entropy, whilst textual entropy averaged about 5\%. When we raise the factor to 3 (i.e., extreme outliers), the numbers drop to about 4.6\% and 3.2\%, respectively. For the purposes of \textbf{R.Q.4}, the relatively high prevalence of outliers in almost all of the projects may provide ways to use entropy as a proxy of the information overhead of introducing new features, as well as a way to measure the spread of information during periods where the information content per file count stabilises or decreases, thus providing a way to quantify the impact of refactoring efforts.
By performing manual inspection of the most extreme events in the sample, we could observe commits where entire new modules were imported or added to the repository, or the deletion of entire sets of files. A more precise characterisation of these events, and how they can contribute to a definition of surprisal, is in the author's agendas.

\section{Limitations And Threats To Validity}
 Although the level of the file is a natural frame of reference for measuring information content of source code, other levels of abstractions - class, module, package, entire project - can also assist in managing complexity. For example, it may be useful to measure the information content of one context relative to another (e.g., file versus package, or package versus project), since this could inform which portions of the code base needs more attention.
 Due to nature of version control systems, precise caching of the histograms that represent different contexts presents a challenge that requires further design and engineering efforts. Finally, the performance of the current process may not be satisfactory in practice, especially for large projects. Although most of the execution times produced an analysis of the entire history in a few hours, execution for big projects with high commit frequency could take 4 or 5 days. As mentioned above, a few of the projects crashed during the execution, sometimes due to failures in PyDriller, and in one occastion the JavaParser virtual machine ran out of memory. These challenges of engineering nature can be worked with by using more powerful machines, as well as by adding defensive code that does not lose the entire execution if processing fails during a specific commit.

\section{Conclusions And Future Work}
In this work, we performed an initial investigation on the entropy of representations of source code in two different main forms. For all the 25 projects, we observed that the correlation between textual and structural entropy is strong (n = 17) or moderate (n = 8). In four of these projects, the tokenisation method was responsible turning a strong correlation into a moderate one. The correlations between structural entropy and number of lines changed, number of tokens changed, and cyclomatic complexity are very weak. For textual entropy, however, 48\% (n = 12) projects showed weak correlations, 12 showed moderate, and a single project displayed high correlation between textual entropy and the three aforementioned metrics.
Although further research on the outliers must be performed before arriving at a more precise conclusion, our counting of both outliers and extremes suggests that extremes might be a better proxy for surprise, given that 12\% of the commits are outliers for structural entropy. Such outliers could assist us in detecting suprising events, which may indicate changes that introduce bugs or big sets of new functionality, as well as efforts to prevent engineers from being overwhelmed by information when dealing with complex code bases.
Further research is also necessary to determine why structural entropy displays such high frequency of outliers.
Future venues of research may also involve separating the tokens of comments from the actual source code. Given that structural entropy was observed to always be smaller than textual entropy, measuring the contribution of comments separately may allow us to know whether comments are useful (or reduntant) to their enclosing context. Such attempts to mange complexity via information theory, if successful, could provide novel tools that could signal the introduction of changes that may be problematic, either becaue the introduce unwanted behaviour, but also because they introduce unwanted complexity, whose long term costs can be just as detrimental of those of bugs.
\bibliographystyle{IEEEtran}
\bibliography{references}

\begin{thebibliography}{10}
\providecommand{\url}[1]{#1}
\csname url@samestyle\endcsname
\providecommand{\newblock}{\relax}
\providecommand{\bibinfo}[2]{#2}
\providecommand{\BIBentrySTDinterwordspacing}{\spaceskip=0pt\relax}
\providecommand{\BIBentryALTinterwordstretchfactor}{4}
\providecommand{\BIBentryALTinterwordspacing}{\spaceskip=\fontdimen2\font plus
\BIBentryALTinterwordstretchfactor\fontdimen3\font minus
  \fontdimen4\font\relax}
\providecommand{\BIBforeignlanguage}[2]{{%
\expandafter\ifx\csname l@#1\endcsname\relax
\typeout{** WARNING: IEEEtran.bst: No hyphenation pattern has been}%
\typeout{** loaded for the language `#1'. Using the pattern for}%
\typeout{** the default language instead.}%
\else
\language=\csname l@#1\endcsname
\fi
#2}}
\providecommand{\BIBdecl}{\relax}
\BIBdecl

\bibitem{McConnell:2004:CCS:1096143}
S.~McConnell, \emph{Code Complete, Second Edition}.\hskip 1em plus 0.5em minus
  0.4em\relax Redmond, WA, USA: Microsoft Press, 2004.

\bibitem{gamma1995patterns}
E.~Gamma, R.~Helm, R.~Johnson, and J.~Vlissides, \emph{Design Patterns:
  Elements of Reusable Object-Oriented Software}.\hskip 1em plus 0.5em minus
  0.4em\relax USA: Addison-Wesley Longman Publishing Co., Inc., 1995.

\bibitem{mcconnel2018techdebt}
S.~McConnell, ``Managing technical debt,'' Available at
  \url{http://www.construx.com/uploadedfiles/resources/whitepapers/Managing\%20Technical\%20Debt.pdf}.

\bibitem{kalyuga2011cognitive}
S.~Kalyuga, ``Cognitive load theory: How many types of load does it really
  need?'' \emph{Educational Psychology Review}, vol.~23, no.~1, pp. 1--19,
  2011.

\bibitem{shannon1948theory}
C.~E. Shannon, ``A mathematical theory of communication,'' \emph{The Bell
  system technical journal}, vol.~27, no.~3, pp. 379--423, 1948.

\bibitem{DEHMER201157history}
\BIBentryALTinterwordspacing
M.~Dehmer and A.~Mowshowitz, ``A history of graph entropy measures,''
  \emph{Information Sciences}, vol. 181, no.~1, pp. 57--78, 2011. [Online].
  Available:
  \url{https://www.sciencedirect.com/science/article/pii/S0020025510004147}
\BIBentrySTDinterwordspacing

\bibitem{dehmer2017mathematical}
\BIBentryALTinterwordspacing
M.~Dehmer, F.~Emmert-Streib, Z.~Chen, X.~Li, and Y.~Shi, \emph{Mathematical
  Foundations and Applications of Graph Entropy}, ser. Quantitative and Network
  Biology.\hskip 1em plus 0.5em minus 0.4em\relax Wiley, 2017. [Online].
  Available: \url{https://books.google.com.au/books?id=5ZMgCwAAQBAJ}
\BIBentrySTDinterwordspacing

\bibitem{ephremides1998networks}
A.~Ephremides and B.~Hajek, ``Information theory and communication networks: an
  unconsummated union,'' \emph{IEEE Transactions on Information Theory},
  vol.~44, no.~6, pp. 2416--2434, 1998.

\bibitem{Konstantinova2011chemistry}
E.~Konstantinova, \emph{Information-Theoretic Methods in Chemical Graph
  Theory}.\hskip 1em plus 0.5em minus 0.4em\relax Boston, MA: Birkh{\"a}user
  Boston, 2011.

\bibitem{Huang536318}
\BIBentryALTinterwordspacing
C.-H. Huang, J.~J.~P. Tsai, N.~Kurubanjerdjit, and K.-L. Ng, ``Computational
  analysis of molecular networks using spectral graph theory, complexity
  measures and information theory,'' \emph{bioRxiv}, 2019. [Online]. Available:
  \url{https://www.biorxiv.org/content/early/2019/04/05/536318}
\BIBentrySTDinterwordspacing

\bibitem{javaSqrtDoc}
O.~Corporation, ``Class strictmath,'' Available at
  \url{https://docs.oracle.com/javase/8/docs/api/java/lang/StrictMath.html}.

\bibitem{soJavaSqrt}
S.~Overflow, ``Where can i find the source code for java's square root
  function?'' Available at
  \url{https://stackoverflow.com/questions/825221/where-can-i-find-the-source-code-for-javas-square-root-function}.

\bibitem{Parr10language}
\BIBentryALTinterwordspacing
T.~Parr, \emph{Language Implementation Patterns: Techniques for Implementing
  Domain-Specific Languages}.\hskip 1em plus 0.5em minus 0.4em\relax Raleigh,
  NC: Pragmatic Bookshelf, 2010. [Online]. Available:
  \url{https://www.safaribooksonline.com/library/view/language-implementation-patterns/9781680500097/}
\BIBentrySTDinterwordspacing

\bibitem{hellerman1972lookup}
L.~Hellerman, ``A measure of computational work,'' \emph{IEEE Transactions on
  Computers}, vol. C-21, no.~5, pp. 439--446, 1972.

\bibitem{Khoshgoftaar1994ApplicationsOI}
T.~M. Khoshgoftaar and E.~B. Allen, ``Applications of information theory to
  software engineering measurement,'' \emph{Software Quality Journal}, vol.~3,
  pp. 79--103, 1994.

\bibitem{davi88complexity}
J.~Davis and R.~LeBlanc, ``A study of the applicability of complexity
  measures,'' \emph{IEEE Transactions on Software Engineering}, vol.~14, no.~9,
  pp. 1366--1372, 1988.

\bibitem{chen1978complexity}
E.~Chen, ``Program complexity and programmer productivity,'' \emph{IEEE
  Transactions on Software Engineering}, vol. SE-4, no.~3, pp. 187--194, 1978.

\bibitem{khoshgoftaar1998information}
T.~M. Khoshgoftaar and E.~B. Allen, ``An information theoretic approach to
  predicting software faults,'' \emph{International Journal of Reliability,
  Quality and Safety Engineering}, vol.~5, no.~03, pp. 227--248, 1998.

\bibitem{berlinger1980complexity}
\BIBentryALTinterwordspacing
E.~Berlinger, ``An information theory based complexity measure,'' in
  \emph{Proceedings of the May 19-22, 1980, National Computer Conference}, ser.
  AFIPS '80.\hskip 1em plus 0.5em minus 0.4em\relax New York, NY, USA:
  Association for Computing Machinery, 1980, p. 773–779. [Online]. Available:
  \url{https://doi.org/10.1145/1500518.1500651}
\BIBentrySTDinterwordspacing

\bibitem{halstead1977elements}
M.~H. Halstead, \emph{Elements of Software Science (Operating and Programming
  Systems Series)}.\hskip 1em plus 0.5em minus 0.4em\relax USA: Elsevier
  Science Inc., 1977.

\bibitem{cook1993information}
C.~Cook, ``Information theory metric for assembly language,'' \emph{Software
  Engineering Strategies}, pp. 52--60, 1993.

\bibitem{mccabe1976complexity}
T.~McCabe, ``A complexity measure,'' \emph{IEEE Transactions on Software
  Engineering}, vol. SE-2, no.~4, pp. 308--320, 1976.

\bibitem{khoshgoftaar1992measure}
T.~Khoshgoftaar and J.~Munson, ``A measure of software system complexity and
  its relationship to faults,'' in \emph{Proceedings of the 1992 International
  Simulation Technology Conference}.\hskip 1em plus 0.5em minus 0.4em\relax The
  Society for Computer Simulation San Diego, CA, 1992, pp. 267--272.

\bibitem{kolmogorov1965three}
A.~N. Kolmogorov, ``Three approaches to the quantitative definition
  ofinformation','' \emph{Problems of information transmission}, vol.~1, no.~1,
  pp. 1--7, 1965.

\bibitem{hassan2009predicting}
A.~E. Hassan, ``Predicting faults using the complexity of code changes,'' in
  \emph{2009 IEEE 31st international conference on software engineering}.\hskip
  1em plus 0.5em minus 0.4em\relax IEEE, 2009, pp. 78--88.

\bibitem{canfora2014changes}
G.~Canfora, L.~Cerulo, M.~Cimitile, and M.~Di~Penta, ``How changes affect
  software entropy: an empirical study,'' \emph{Empirical Software
  Engineering}, vol.~19, no.~1, pp. 1--38, 2014.

\bibitem{keenan2022investigation}
D.~Keenan, D.~Greer, and D.~Cutting, ``An investigation of entropy and
  refactoring in software evolution,'' in \emph{International Conference on
  Product-Focused Software Process Improvement}.\hskip 1em plus 0.5em minus
  0.4em\relax Springer, 2022, pp. 282--297.

\bibitem{allamanis2016convolutional}
M.~Allamanis, H.~Peng, and C.~Sutton, ``A convolutional attention network for
  extreme summarization of source code,'' in \emph{International conference on
  machine learning}.\hskip 1em plus 0.5em minus 0.4em\relax PMLR, 2016, pp.
  2091--2100.

\bibitem{movshovitz-attias-cohen-2013-natural}
\BIBentryALTinterwordspacing
D.~Movshovitz-Attias and W.~W. Cohen, ``Natural language models for predicting
  programming comments,'' in \emph{Proceedings of the 51st Annual Meeting of
  the Association for Computational Linguistics (Volume 2: Short
  Papers)}.\hskip 1em plus 0.5em minus 0.4em\relax Sofia, Bulgaria: Association
  for Computational Linguistics, Aug. 2013, pp. 35--40. [Online]. Available:
  \url{https://aclanthology.org/P13-2007}
\BIBentrySTDinterwordspacing

\bibitem{javaparser}
F.~T. Nicholas~Smith, Danny van~Bruggen, ``Javaparser,'' Available at
  \url{https://github.com/javaparser/javaparser}.

\bibitem{Dabic:msr2021githubSearch}
O.~Dabic, E.~Aghajani, and G.~Bavota, ``Sampling projects in github for {MSR}
  studies,'' in \emph{18th {IEEE/ACM} International Conference on Mining
  Software Repositories, {MSR} 2021}.\hskip 1em plus 0.5em minus 0.4em\relax
  {IEEE}, 2021, pp. 560--564.

\bibitem{munaiah2017curating}
\BIBentryALTinterwordspacing
N.~Munaiah, S.~Kroh, C.~Cabrey, and M.~Nagappan, ``Curating github for
  engineered software projects,'' \emph{Empirical Softw. Engg.}, vol.~22,
  no.~6, p. 3219–3253, dec 2017. [Online]. Available:
  \url{https://doi.org/10.1007/s10664-017-9512-6}
\BIBentrySTDinterwordspacing

\bibitem{PyDriller}
D.~Spadini, M.~Aniche, and A.~Bacchelli, ``Pydriller: Python framework for
  mining software repositories,'' in \emph{The 26th ACM Joint European Software
  Engineering Conference and Symposium on the Foundations of Software
  Engineering (ESEC/FSE)}, 2018.

\bibitem{Hucka2018spiral}
\BIBentryALTinterwordspacing
M.~Hucka, ``Spiral: splitters for identifiers in source code files,''
  \emph{Journal of Open Source Software}, vol.~3, no.~24, p. 653, 2018.
  [Online]. Available: \url{https://doi.org/10.21105/joss.00653}
\BIBentrySTDinterwordspacing

\bibitem{anonymousRepo}
``https://anonymous.4open.science/r/nlbse2023entropy-d6e9/,'' Available at
  \url{https://anonymous.4open.science/r/nlbse2023entropy-D6E9/}.

\bibitem{lehman1980laws}
M.~Lehman, ``Programs, life cycles, and laws of software evolution,''
  \emph{Proceedings of the IEEE}, vol.~68, no.~9, pp. 1060--1076, 1980.

\bibitem{akoglu2018coefficients}
H.~Akoglu, ``User's guide to correlation coefficients,'' \emph{Turkish Journal
  of Emergency Medicine}, vol.~18, 08 2018.

\end{thebibliography}
\end{document}

\end{document}